\documentclass[aps,floats,superscriptaddress,twocolumn,pre,showpacs]{revtex4-1}

\usepackage{graphicx,epsfig}
\usepackage{graphics,dcolumn,bm,epic, eepic,fleqn,float}
\usepackage{amssymb,amsmath,amsfonts,multirow,rotate,color}
\usepackage{soul}
\usepackage{wasysym}
\usepackage{dcolumn}
\usepackage{bm}
\usepackage{hyperref}

\definecolor{Myorange}{cmyk}{0,0.42,1,0}

\newcommand{\lay}[1]{^{[#1]}}

\begin{document}

\title{Multilayer motif analysis of brain networks}

\author {Federico Battiston}
\affiliation{School of Mathematical Sciences, Queen Mary University of London, 
London E1 4NS, United Kingdom}   

\author {Vincenzo Nicosia}
\affiliation{School of Mathematical Sciences, Queen Mary University of London, 
London E1 4NS, United Kingdom}  

\author {Mario Chavez}
\affiliation{CNRS-UMR-7225, H\^opital Piti\'e Salp\^etri\`ere, 75013 Paris, France}   

\author{Vito Latora}
\affiliation{School of Mathematical Sciences, Queen Mary University of London, 
London E1 4NS, United Kingdom}  

\affiliation{Dipartimento di Fisica ed Astronomia, Universit\`a di Catania and INFN, I-95123 Catania, Italy}

\date{\today}

\begin{center}
\begin{abstract}
In the last decade, network science has shed new light both on the
structural (anatomical) and on the functional (correlations in the
activity) connectivity among the different areas of the human
brain. The analysis of brain networks has made possible to detect the
central areas of a neural system, and to identify its building blocks
by looking at overabundant small subgraphs, known as motifs. However,
network analysis of the brain has so far mainly focused on anatomical
and functional networks as separate entities. The recently developed
mathematical framework of multi-layer networks allows to perform an
analysis of the human brain where the structural and functional layers
are considered together. In this work we describe how to classify the
subgraphs of a multiplex network, and we extend motif analysis to
networks with an arbitrary number of layers. We then extract
multi-layer motifs in brain networks of healthy subjects by
considering networks with two layers, anatomical and functional,
respectively obtained from diffusion and functional magnetic resonance
imaging. Results indicate that subgraphs in which the presence of a
physical connection between brain areas (links at the structural
layer) coexists with a non-trivial positive correlation in their
activities are statistically overabundant. Finally, we investigate the
existence of a reinforcement mechanism between the two layers by
looking at how the probability to find a link in one layer depends on
the intensity of the connection in the other one. Showing that
functional connectivity is non-trivially constrained by the underlying
anatomical network, our work contributes to a better understanding of
the interplay between structure and function in the human brain.
\end{abstract}
\end{center}

\maketitle

\textbf{Many networks are characterised by the presence of non-trivial structures at the microscopic scale. In particular, biological networks are rich in certain subgraphs, because these are crucial for the stability of the system and for the efficient processing of information. Motifs have been largely studied in neuroscience both in networks of anatomical connectivity and in networks of correlations in the 
functional activity of different brain regions. To shed new lights on the intimate relations between  structure and function in the human brain, we consider the two networks as the layers of a multiplex brain network, and we investigate the presence of statistically overabundant subgraphs spanning across the two layers in several healthy subjects. We provide a mathematical framework for multi-layer motif analysis and identify over-represented subgraphs associated to the existence of overlap and structural balance in the two layers, as well as the existence of significant reinforcement mechanisms among the structural and functional connections in the human brain.}

\section{Introduction}

From the brain to the Internet and to social groups, the characterization of the connectivity patterns of complex systems has revealed a wiring organization that can be captured neither by regular lattices, nor by random graphs~\cite{boccaletti2006}. In neurosciences, it is widely acknowledged that the emergence of several pathological states is accompanied by alterations in brain connectivity patterns~\cite{stam2014, fornito2014}. 
Indeed, empirical studies have lead to the hypothesis that the brain 
relies on the coordination of a scattered mosaic of distant brain regions, forming non-random 
a weblike structure of neural assemblies, and that brain dysfunctions are related to a lack 
of such coordination ~\cite{varela01}.  

In the last decade, the combined use of advanced neuroimaging
tecniques and of mathematical tools to characterise the structure of a
complex network has significantly improved our understanding of how
the brain works. However, it is important to notice that we have two
fundamentally different ways to study brain connectivity, since data
can reflect either anatomical properties of the brain, or functional
neural activity.

In the first case ({\em anatomical brain networks}) we construct
networks whose nodes are usually putative brain regions and the links
represent physical connections among them, while in the second case
({\em functional brain networks}) each node represents an area of the
brain, usually consisting of neural assemblies, and each link
indicates the presence of a functional interaction between the
activity (electrical, magnetic or hemodynamic/metabolic) of two areas.

Despite the evident relations between the two type of networks, 
comparison between anatomical and functional brain networks is not straightforward~\cite{deco2011, nicosia2014}.  Theoretical studies support the idea that anatomical connections can determine some aspects of brain dynamics~\cite{deco2011}, but it is less clear how in general the anatomical connectivity supports or facilitates 
the emergence of the properties of functional networks. 

Correspondence between functional and structural networks remains thus an active research area~\cite{honney2007, honney2009, honney2010}. A better understanding of how anatomical connections support  communication, correlations and synchronisation of brain activities is necessary to read normal neural processes, as well as to improve the identification and prediction of alterations in brain diseases.

In this work we contribute to unveiling the delicate relations between
structure and function in the human brain by focusing on {\em network
  motif analysis}, a tool that has revealed quite successful in
network science.  A network motif is a small subgraph that is
statistically over-represented in a complex network with respect to a
given null model~\cite{milo02}. Empirical evidence suggests that
motifs are a key concept from RNA structures to social
networks~\cite{milo02, mangan03}. The emergence of motifs, i.e. the
abundance of certain types of subgraphs in a given network, seems to
be related to the robustness of the system, or to the stability of the
dynamical or signaling circuits that each motif
represents~\cite{milo02, mangan03}.

In previous studies, functional interactions have been found to variate with the patterns of local structural motifs in the monkey cortex~\cite{adachi2012}. Similarly, functional integration of cortical areas in monkeys seem to be strongly determined by some properties (e.g. density and symmetry) of structural motifs~\cite{shen2012}. Neurocomputational modeling indicates, for instance, that neuronal  networks motifs might play a role on information transmission delays and on long- and short-term memory~\cite{li2008}. Recent results suggest that network motifs analysis can provide significant new markers for the progression of Alzheimer's disease~\cite{friedman2015}. Motif analysis has been applied separately, both to anatomical and functional brain and, although some motifs which are considered central to information processing in the brain have emerged~\cite{sporns04, sakata2005, iturria2008},  the interplay between structural and functional motifs is still not well understood.  

Here, we investigate the relation between structure and function in the brain by generalising  
motif analysis to the case of multiplex networks, and by detecting {\em multiplex motifs} in the 
brain.  The concept of multi-layer networks has been recently introduced in network theory to deal with systems whose nodes are coupled through different types of interactions~\cite{boccaletti2014, kivela14, battiston16challenges, dedomenico2013, battiston14}. This novel formalism has been recently applied to connectivity matrices estimated from magnetoencephalographic data for getting a more complete picture of neural interaction across different frequency bands~\cite{brookes2016, dedomenico16}.

The multiplex motifs we are interested in are small multi-layer
connected subgraphs which are statistically overabundant in multiplex
networks describing real systems. The related problem of finding
isomorphisms in multi-layer networks has been considered in
Ref.~\cite{kivela15}. The layered organization of triadic connections
and clustering has been studied in Refs.~\cite{battiston14, cozzo15,
  barrett12, brodka12, criado11}.  The non-trivial overlap across the
layers of larger mesoscale structures, such as communities, found in
multiplex networks from the real world has been investigated in
Ref.~\cite{mucha10, dedomenico15, battiston16}. In this paper we use
multi-layer motif analysis to study multiplex networks with two layers
constructed from structural and functional information on the brains
of healthy subjects, respectively obtained by Diffusion Magnetic
Resonance Imaging (DW-MRI) and resting-state functional MRI
(rs-fMRI). In these networks, nodes are defined as Regions of Interest
of the brain (ROIs). The edges of the structural DW-MRI layer
represent the estimated white matter connection strength between any
pair of ROIs, while links in the functional network indicate
functional correlations between the fMRI time-series of the two
corresponding ROIs.  As we will show, our approach to detect multiplex
anatomical/functional motifs is able to provide useful insights on
multiplex networks derived from multiple brain modalities of healthy
subjects, as well as multimodal connectivity alterations due to
various brain disorders.

The paper is organised in the following way.  We first present in
Section II the mathematical framework needed to define motifs in
the context of multiplex networks. In Section III we describe the
dataset of multilayer brain networks that we analyse in Section IV.
In Section V we investigate the existence of network reinforcement
mechanisms in the human brain, and in Section VI we provide a critical
discussion of the results and some conclusions.

\section{Multilayer motif analysis}

\begin{figure*}[!ht]
\begin{center}
\includegraphics[width=6in]{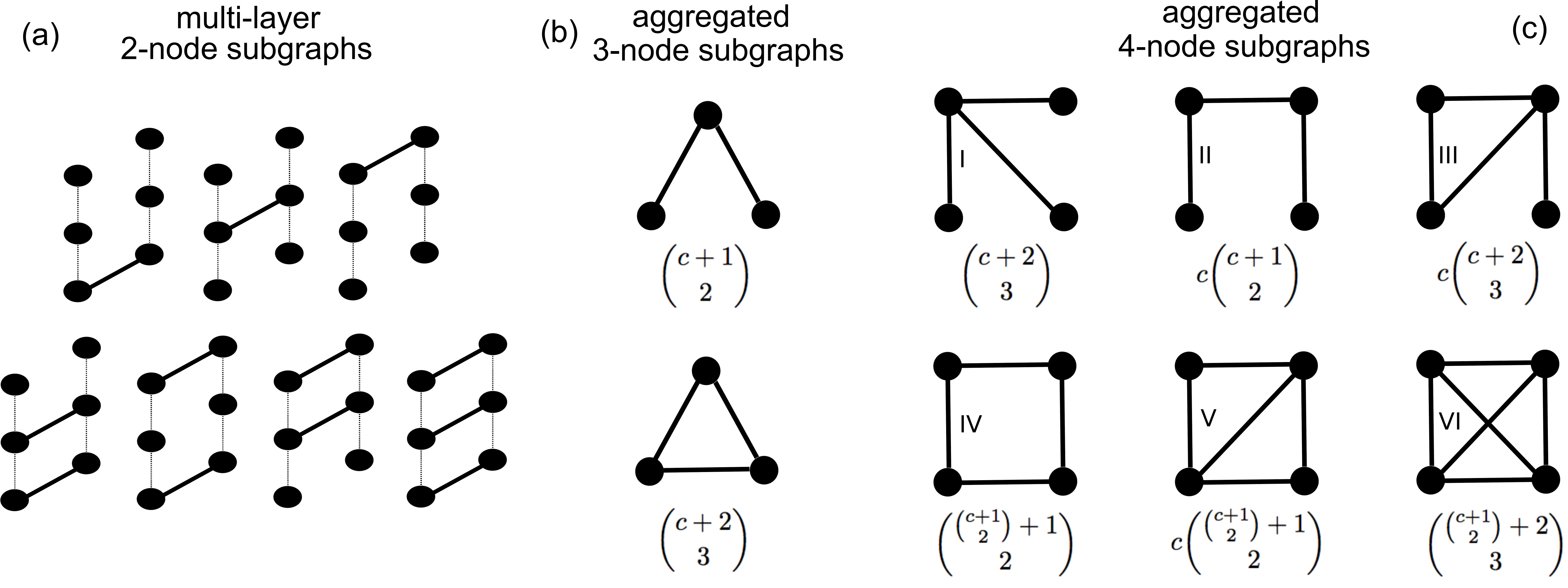}
\caption[]{A simple edge on the aggregated network can originate from
  different multi-layer subgraphs. For $M=3$ layers there exist $c=7$
  different types of multi-links ($s=1$, see Eq.~\ref{eq:2}) (a). At
  the aggregated level, we distinguish two subgraphs with $n=3$ nodes
  (b), the chain (top) and the clique (bottom), and six motifs with
  $n=4$ nodes (c), usually respectively known as the star, the chain
  and the 3-loop-out (top, left to right), the box, the semi-clique
  and the clique (bottom, left to right). We report the number of
  different multi-layer motifs giving rise to each aggregated subgraph
  as a function of the number of multi-links $c$.}
\label{fig1bis}
\end{center}
\end{figure*}

\textbf{Single-layer motifs.}  In single-layer networks, standard
motif analysis searches for small subgraphs that are statistically
over-represented in a given graph $G$ with respect to a null
model~\cite{milo02}. In practice, the frequency of each subgraph $g$
in graph $G$ is compared with the expected frequency of that subgraph
in an appropriately randomised version of the graph $G$, e.g. in the
family of random graphs having the same number of nodes and the same
number of edges of $G$. If the actual frequency of the subgraph $g$ in
in $G$ is larger than that expected in the null model and the
difference is statistically significant, then $g$ is an
overrepresented subgraph, i.e. a network motif.  Small connected
subgraphs are typically classified at two different levels: they are
first sorted according to their number of nodes $n$ and then
classified based on the number $\ell$ and placement of their links.
If $G$ is undirected, the smallest subgraphs of interest are those
with $n=3$ nodes. In this case, two different connected subgraphs can
be identified, namely the chain, also known as triad, and the complete
graph, or triangle.  We have then six different subgraphs with $n=4$
nodes, nineteen subgraphs with $n=5$ nodes, with this number growing
fast as the number of nodes increases. Because of the decreasing
statistical significance of larger subgraphs and the growing
computational cost associated to their detection, motif analysis in
real-world networks is usually limited to small subgraphs consisting
of a few nodes.

\textbf{Motifs in multi-layer networks.} A multiplex network $\mathcal
M$ is a convenient representation for a system in which the nodes are
related through different types of interactions, which can be
represented as layers. If a multiplex network has $M$ types of
connections, it will consist of M distinct layers, and can be
represented by the vector of adjacency
matrices~\cite{battiston14,bianconi13}
\begin{equation}
{\mathcal M}=\{A^{[1]}, ... , A^{[M]}\}.
\end{equation}
where $A^{[\alpha]}$ is the adjacency matrix of layer $\alpha$. In the
case of binary interactions $A^{[\alpha]}=\{a_{ij}^{[\alpha]}\}$, with
$a_{ij}^{[\alpha]}=1$ if nodes $i$ and $j$ are connected on layer
$\alpha$, and $a_{ij}^{[\alpha]}=0$ otherwise.  When the nature of the
links is neglected, the system is described by the corresponding
aggregated network $\mathcal A = \{a_{ij}\}$, where
\begin{equation}
 a_{ij} = \begin{cases}1 & \text{if} \>\>\>\exists\> \alpha:
   a\lay{\alpha}_{ij}=1 \\ 0 & \text{otherwise.} \end{cases}
\end{equation} 

The simplest approach would be then a single-layer motif analysis on
network $\mathcal A$. While such analysis provides useful information
on micro-scale connectivity patterns at the aggregated level,
information is lost on the significance of the layered organisation of
interactions.  A given subgraph $g$ observed in the aggregated graph
$\mathcal A$ can indeed originate from different combinations of the
edges across the layers of the system, so that we are interested to
enumerate the different types of multi-layer motifs contributing to
each subgraph in the aggregate network.  We suggest here a
classification of motifs in multiplex networks on three levels.  At
the \emph{first level}, multi-layer connected subgraphs are categorised according to
their number of nodes $n$. At the \emph{second level}, they are classified according
to the subgraphs they generate in the corresponding aggregated network
$\mathcal A$. At the \emph{third level}, for each subgraph in $\mathcal A$, the
different multiplex subgraphs are distinguished by looking at the
exact pattern of connections across the different layers.

\begin{figure*}[!ht]
\begin{center}
\includegraphics[width=6.5in]{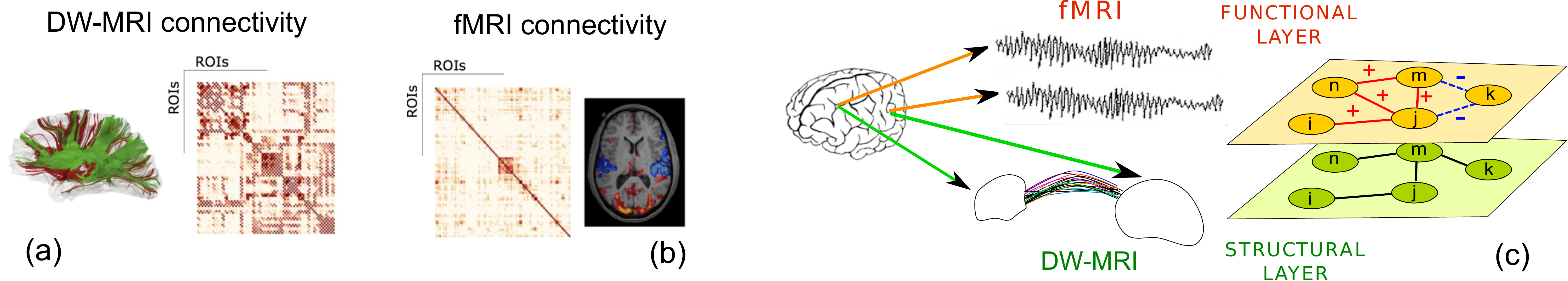}
\caption[]{DW-MRI connectivity provides the strength of axonal connections among regions of interest (ROIs) of the brain for the anatomical network (a). Positive and negative correlations in the activity of the different areas are measured through fMRI for the functional layer (b). Such information can be combined to construct a multi-layer brain network (c).}
\label{fig1}
\end{center}
\end{figure*}

\textbf{Multi-layer motifs with $n=2$ nodes.} Due to the richness
provided by the presence of more than one layer, the smallest motifs
of interest in multiplex networks are already those with $n=2$
nodes. Such multilayer motifs correspond in practice to the different
types of \emph{multi-links} connecting two nodes~\cite{bianconi13}. An edge
between two nodes appears in the aggregated network if the two nodes
are connected on at least one of the layers. We now derive the total
number of different multi-layer configurations giving rise to a single
link in $\mathcal A$. In particular, in a multiplex with $M$ layers we
have ${M \choose 1}$ configurations where an edge between two nodes
exists only in one of the layers, ${M \choose 2}$ configurations such
that two nodes are linked at two different layers, and so on. Hence,
the total number of distinct multi-links is equal to
\begin{equation}
c =\sum_{m=1}^M {M \choose m} = 2^M - 1,
\label{eq:1}
\end{equation}
where we have neglected the degenerate configuration in which there is
no edge between the two nodes at any layer, since in that case the
corresponding subgraph is disconnected. Let us consider for instance a multiplex network with $M=3$ layers. As shown in Figure~{\ref{fig1bis}(a)}, we have 
in total $c=7$ possible motifs with $n=2$ nodes, of which ${3 \choose 1}=3$ have one edge at a single layer, ${3 \choose 2}=3$ with edges on two layers and one complete multi-link with edges at all levels $\biggl({3 \choose 3}=1\biggr)$.

An interesting case is that where the edges at a given layer can be of
different types, e.g. signed or coloured edges, and such information
is lost in the aggregate graph. Let us denote by $s$ the number of
different types of edges allowed on each layer (for instance $s=2$ for
a signed network with positive and negative edges), with $s$ equal on
every layer. In such case, the total number of multi-links is equal to
\begin{equation}
c =\sum_{m=1}^M s^m {M \choose m} = (s+1)^M - 1.
\label{eq:2}
\end{equation}

If each layer $\alpha$ has a different number $s^{[\alpha]}$ of edge
types, the total number of multi-links is equal to
\begin{equation}
c =\biggr[  \prod_{\alpha=1}^M (s^{[\alpha]}+1) \biggl] - 1.
\label{eq:3}
\end{equation}
The latter formula correctly reduces to Eq.~(\ref{eq:2}) and
Eq.~(\ref{eq:1}) respectively if $s^{[\alpha]}=s$ $\forall \alpha$, and
if $s=1$.

\textbf{Multi-layer motifs with $n=3$ nodes.} For motifs with more
than $n=2$ nodes, the problem of counting the number of multi-layer
configurations of a given aggregated motif is equivalent to that of
finding the number of non-isomorphic ways in which that subgraph can
be coloured by choosing edges from $c$ different colours. 

Let us focus
first on the case of connected subgraphs with $n=3$ nodes. In such
case we distinguish two different subgraphs on the aggregated network
$\mathcal A$, the triad $g_{n=3,\ell=2}$ and the triangle
$g_{n=3,\ell=3}$, shown respectively at the top and the bottom of Figure~\ref{fig1bis}(b). Let us consider a \emph{triad} formed by $\ell=2$ specific
multi-links, each chosen from the $c$ possible ones. In general there are $c^2$ ways to colour a labeled triad. However, each possible configuration of two coloured multi-edges generates $2$ isomorphic configurations. Hence, the number $t$ of different
multi-layer triads is equal to the number of different unordered
selections with repetition of $\ell=2$ multi-links of $c$ types, i.e.
\begin{equation}
t={c + 1 \choose 2}
\label{eq:triads}
\end{equation}
Similarly, let us consider a \emph{triangle} formed by $\ell=3$ specific multi-links. In general there are $c^3$ to colour a labeled triangle. However, each possible configuration of two coloured multi-edges generates $3!=6$ isomorphic configurations. Consequently, the number $T$ of multi-triangles is equal to the different unordered selections with repetition of $\ell=3$ multi-links, i.e.
\begin{equation}
T={c + 2 \choose 3}
\label{eq:triangles}
\end{equation}

\begin{figure*}[!t]
\begin{center}
\includegraphics[width=6in]{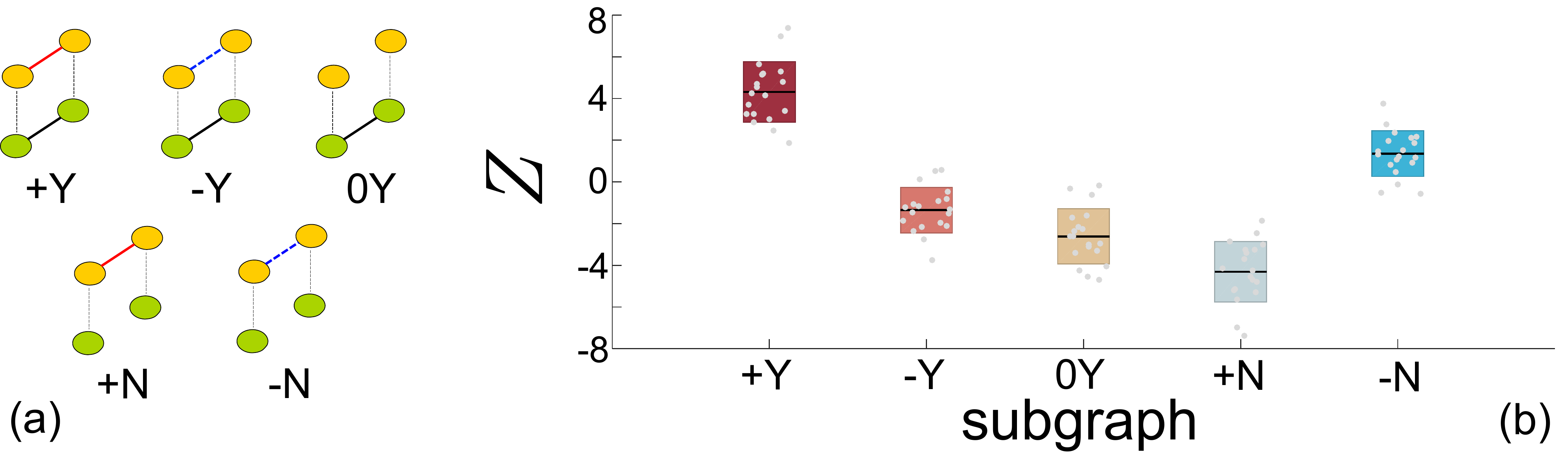}
\caption[]{The presence (Y) or absence (N) of links in the structural layer $\mathcal D$, and the existence of a significant positive link (+), negative link (-) or its absence (0) in the functional layer $\mathcal F$ give rise to $c=5$ (see Eq.~\ref{eq:3}, $s^{[d]}=1$, $s^{[f]}=2$) (a). The Z-scores of the subgraphs show that +Y is a strongly over-represented motif, indicating correlation between the co-activation of two ROIs and the existence of a structural link. Conversely, the motif -Y is roughly as
likely in real and random data (actually slightly under-represented in real data), suggesting that significant negative functional links appear to be at random relatively to the physical connections (b). Results are shown for 19 healthy subjects.}
\label{fig2a}
\end{center}
\end{figure*}

\textbf{Multi-layer motifs with $n>3$ nodes.} At
difference with the multi-layer triads and triangles, larger multiplex
subgraphs are in general not characterised by symmetry classes as simple as those of the triad or triangle. As a consequence, counting the
number of multi-layer patterns associated to the same aggregated subgraph
is in general more complicated. The basic idea is to decompose each
structure as a combination of smaller motifs, such as multi-links (whose multiplicity
is equal to $c$), pairs and triples consisting of symmetric multi-links (for which the previous formulas introduced in
Eqs. (\ref{eq:triads}) and (\ref{eq:triangles}) apply).

For simplicity, let us focus on the motifs with $n=4$ nodes shown in Figure~\ref{fig1bis}(c). The \emph{star} I is the only motif where all the links are indistinguishable. Hence, it can be generated in a number of multi-layer configurations corresponding again to selecting $\ell=3$ unordered multi-links with repetition. For the \emph{chain} II, the multi-layer multiplicity can be determined as the product between the number of possible central multi-links and that of the pair of symmetric external edges. The \emph{3-loop-out} III corresponds to the product of the number of multi-links with the number of multiplex triangles. For the \emph{box} IV, we can decompose the problem into selecting in an unordered way with repetition two different pairs, each composed of two symmetric multi-links. The multi-layer multiplicity of the \emph{semi-clique} V is equal to the product of one multi-link and a box. At last, for the multi-layer \emph{clique} VI we can decompose the problem into selecting in an unordered way with repetition three different pairs, each composed of two symmetric multi-links. Similar techniques can be used to compute the multi-layer multiplicity of motifs with $n>4$ nodes. The exact number of multi-layer configurations corresponding to the different motifs with $n=3$ or $n=4$ nodes is reported in Figure~\ref{fig1bis}.

\section{Data}

We study a data set of the multiplex brain networks of 19 control subjects obtained from the USC Multimodal Connectivity Database (\url{http://umcd.humanconnectomeproject.org}), an open repository for brain connectivity matrix sharing and analysis~\cite{bronw2016}. Each multiplex consists of two layers representing the structural (anatomical) and functional connections among the brain areas of the corresponding subject, respectively inferred by means of Diffusion Magnetic Resonance Imaging (DW-MRI) and functional Magnetic Resonance Imaging (fMRI).  
The anatomical connectivity network is based on the connectivity matrix obtained by DW-MRI data from 19 healthy participants~\cite{rudie2013}.  Whole brain deterministic tractography was performed using the fiber assignment by continuous tracking algorithm. Fiber tractography was carried out by propagating fibers from each voxel with a maximum turn angle of 50$^{\circ}$ followed by a spline filter smoothing. Each element of the matrix represents thus an approximation of the anatomical strength between the corresponding pair of brain regions. 
The elements of this matrix gives the estimated number of fibers between different anatomical regions 
of interest ($N = 264$ nodes in all the networks)  spanning the cerebral cortex, subcortical structures, and the cerebellum~\cite{power2011}. 

The functional brain connectivity was extracted from fMRI recordings obtained in~\cite{rudie2013}. All fMRI data sets (segments of 6 minutes recorded from the same 19 healthy subjects) were normalized, corrected and sub-sampled from the same set of anatomical regions as for anatomical connectivity. These 264 putative functional regions were shown to more accurately represent the information present in the brain network relative to other voxelwise and atlas-based parcellation approaches~\cite{rudie2013}. To eliminate low frequency noise (e.g. slow scanner drifts) and higher frequency artifacts from cardiac and respiratory oscillations, time-series were digitally filtered with a finite impulse response filter with zero-phase distortion (bandwidth $=0.01 - 0.1$ Hz)~\cite{rudie2013}. For the functional connectivity, linear correlation were estimated between time series of each of the 264 brain regions. Correlation coefficient were then variance-stabilized by applying the Fisher's Z-transform in order to generate $264 \times 264$ whole brain functional connectivity matrices for each subject.

\begin{figure*}[!ht]
\begin{center}
\includegraphics[width=0.9\textwidth]{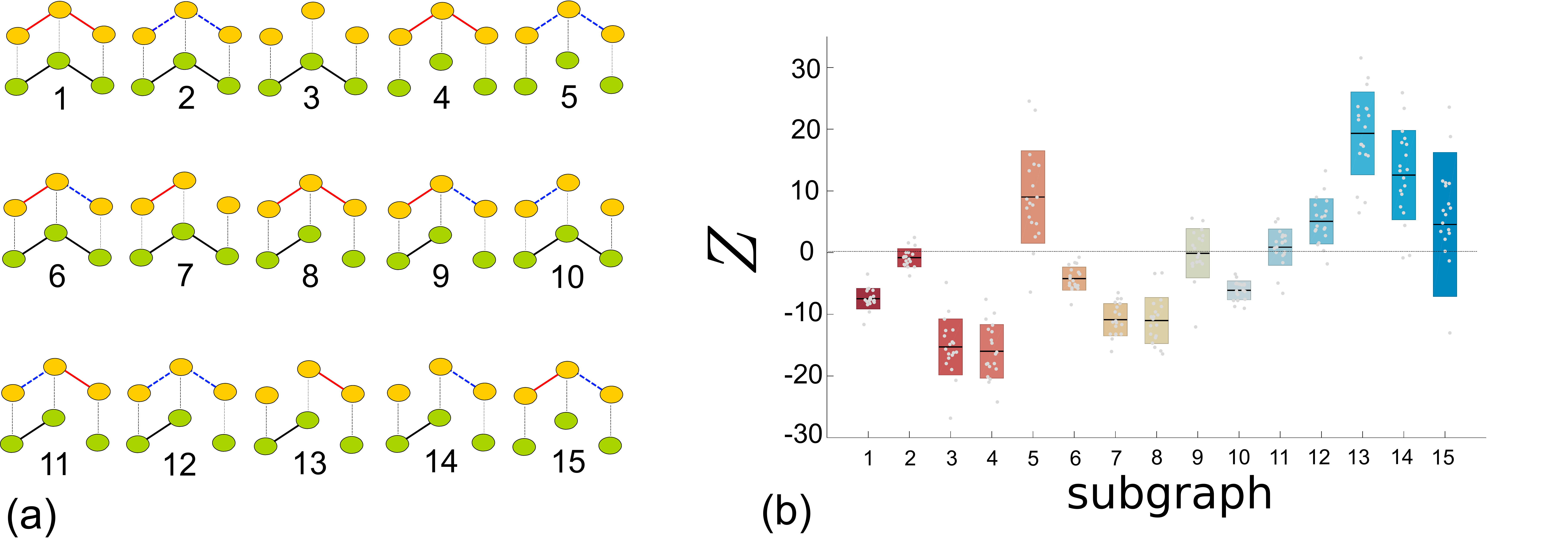}
\caption[]{The $t=15$ different multi-layer triads are shown in (a). Their corresponding value of the Z-score are reported in (b). Triad $5$, $13$ and $14$ emerge as statistically validated recurrent motifs ($p<0.05$, corrected for multiple comparisons).}
\label{fig2b}
\end{center}
\end{figure*}

Consequently, the weight of an edge in the DW-MRI layer indicates the presence (and strength) of axonal connections between the corresponding areas, while the weight of an edge in the fMRI layer is proportional to the correlation in the time-series of hematic flow activity associated to the two areas. In the following we will refer to $\mathcal D=\{d_{ij}\}$ as the physical connectivity matrix, where $d_{ij}$ represents the weight of the physical connection between node $i$ and node $j$. For each subject such matrix consists of a single connected component and it is sparse ($d_{ij}=0$ for many pairs $i,j$.). Functional data, conversely, in principle represent a fully connected graph. For such reason we thresholded such graph, keeping only significant functional links (both positive and negative ones). The significance was set at $p <0.05$, corrected for multiple comparisons. We will refer to such thresholded functional graphs as $\mathcal F=\{f_{ij}\}$, where $f_{ij}$ represents the weight of the significant functional connection between node $i$ and node $j$ (if the link is not significant we set $f_{ij}=0$). We indicate by
\begin{equation}
{\mathcal M}=\{\mathcal D, \mathcal F \}.
\end{equation}
the multiplex brain network $\mathcal M$ encoding information on such structural and functional layers, and illustrated in Figure~\ref{fig1}.
The correlation $f_{ij}$ between the fMRI activity of two ROIs $i$ an
$j$ can be either positive (+, red links), or negative (-, blue links) or non-significative
(0, no link). Conversely a  structural edge between two ROIs might either exist (Y, green links) or
not (N, no link). 

\section{Motifs in multi-layer brain networks}

\begin{figure*}[!ht]
\begin{center}
\includegraphics[width=0.95\textwidth]{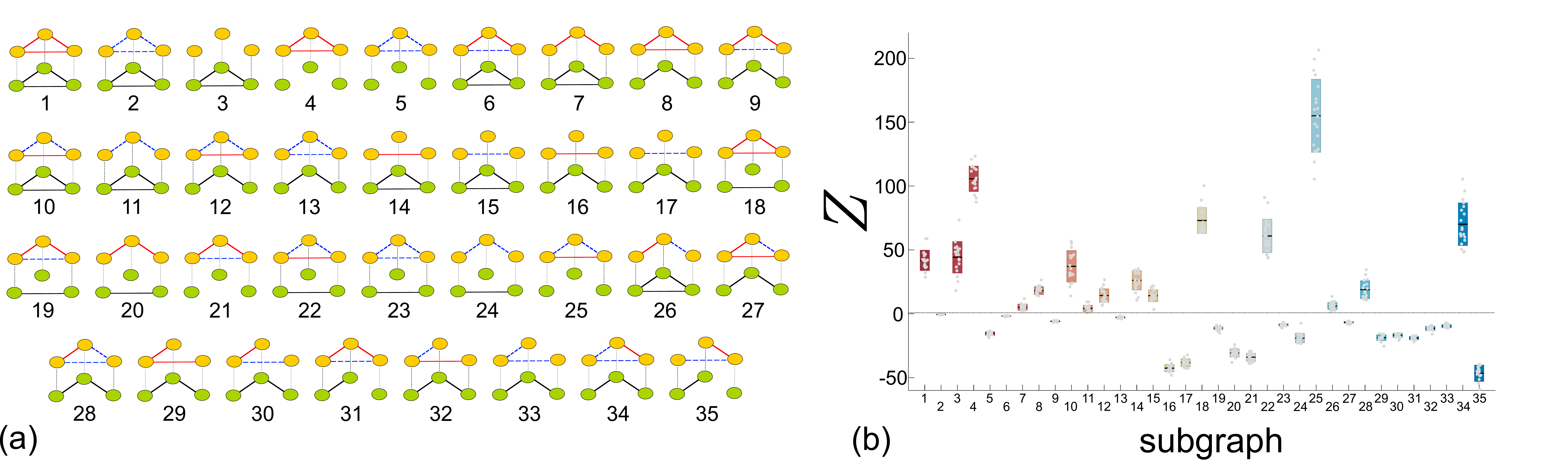}
\caption[]{The $T=35$ different multi-layer triangles are shown in (a). Their corresponding value of the Z-score are reported in (b). Triangles $1$, $4$, $18$, $22$, $25$ and $34$ emerge as statistically validated recurrent motifs.}
\label{fig2c}
\end{center}
\end{figure*}

We now move our attention to the investigation of multi-layer motifs
in brain networks. We first consider the structural layer $\mathcal D$
and the functional layer $\mathcal F$ as binary unweighted networks.
Since links are signed in the functional layer $\mathcal F$,
i.e. $s^{[f]}=2$, and simple in the structural layer $\mathcal D$,
i.e. $s^{[d]}=1$, from Eq.~(\ref{eq:3}) we have $c=5$ elementary motifs
with $n=2$, namely +Y , -Y , 0Y , +N , -N. They are illustrated in
Figure~\ref{fig2a}(a).

We report in Figure~\ref{fig2a}(b) the Z-score of each multi-link $g$
\begin{equation}
Z(\mu_{g})=\frac{\mu_{g}-\bar \mu_{g}}{\sigma_{g}},
\end{equation}
where the abundance $\mu_{g}$ of the subgraph in the real system is compared with the average abundance $\bar \mu_{g}$ and standard deviation $\sigma_{g}$ of what found on a suitable randomised network. A high positive value $Z(\mu_{g})$ indicates that $g$ is a significant recurrent motif.

In order to consider a suitable null-model, it is necessary to take into account the division of the brain into two distinct hemispheres. For such a reason, we randomise the links of the layer by performing the following block configuration model. For each node, tot only we preserve its total degree at that layer, but we also keep fixed its number of connections towards regions in the same brain hemisphere and those towards the other hemisphere. Such block configuration model can be practically implemented by performing two standard configuration models for the intra-hemisphere links in the right and in the left hemisphere, and by performing a bipartite configuration model on the inter-hemisphere links. In Figure~\ref{fig2a}(b) results are shown for a multiplex null-model where we kept fixed the signed functional layer and performed a block configuration model on the structural layer, with 100 randomisation for each subject.

Interestingly, the motif +Y (corresponding to the concurrent presence
of a positive fMRI correlation and of a direct connection in the DTI
layer) is significantly over-represented, while +N (positive
correlation and absence of edge) is markedly underrepresented.
This is a remarkable result as it supports the hypothesis that functional positive links are definitely
correlated with the structural network. Conversely the motif -Y is as
likely in real data as in the random model, which indicates that two brain areas physically connected do not correlate with negative functional interactions between their dynamics. Results are shown for 19 different subjects.

Having fixed the abundance of the $c=5$ multi-links, we are interested in knowing the significance of each higher order motif. It is possible to extend such motifs analysis to larger subgraphs, such as motifs with $n=3$ nodes. According to Eqs.~(\ref{eq:triads}) and ~(\ref{eq:triangles}), in the considered multiplex networks there are $t=15$ multi-layer triads and $T=35$ multi-layer triangles. For each subgraph with $n=3$ we compute now the Z-score by comparing the value found in the real data with the average value and standard deviations of a randomised ensemble of networks, where the unsigned structure of the overlapping network is kept fixed, but the different multi-links were shuffled. In Figure~\ref{fig2b} we see for instance that, even if the multi-link +Y is over-represented, the triad formed by two multi-links of such type is underrepresented, with $Z \approx -8$. This is due to the fact that in the real data, if one region $i$ is connected to two other regions $j$ and $k$ with both a structural and positive functional links, it is very unlikely that $j$ and $k$ are no connected at all.  We notice that for the same reason almost all multi-layer triads with at least one multi-link of type +Y have a negative Z-score, as shown in Figure~\ref{fig2b}, with the exception of triad of type 9. In this subgraph, the central node is indeed strongly connected to one of the other region, both physically and with a positive functional link, but only weekly connected to the other with a negative functional link and no structural link. Hence, the two external regions of the triads do not appear to be communicating much, and the resulting Z-score is approximately 0.

We now analyse which other multi-layer triangles are overabundant in the real data and propose possible underlying reasons for this phenomenon.
The  high Z-score of the first multi-layer triangles in Figure~\ref{fig2c} confirms that,  if one region $i$ is connected to two other regions $j$ and $k$ at both layers, $j$ and $k$ are also directly connected by links at both layers as well. Triangles of type 3 are significative due to high clustering in the structural layer. Triangles of type 4 have a high Z-score because of structural balance (three positive values of correlations) in the functional layer. Similarly, many other multi-layer triangles appear to be overabundant due to the existence of signed balanced triangles in the functional layer (given either by three positive links, as in triangles of type 1, 4 and 18, or by one positive link and two negative link, as for triangles of type 10, 22, 25, 34). The mentioned motifs were confirmed to be statistically over-represented from the reference null model by using a $p<0.05$ significance level.

\section{Network reinforcement mechanisms}

\begin{figure*}[!t]
\begin{center}
\includegraphics[width=0.9\textwidth]{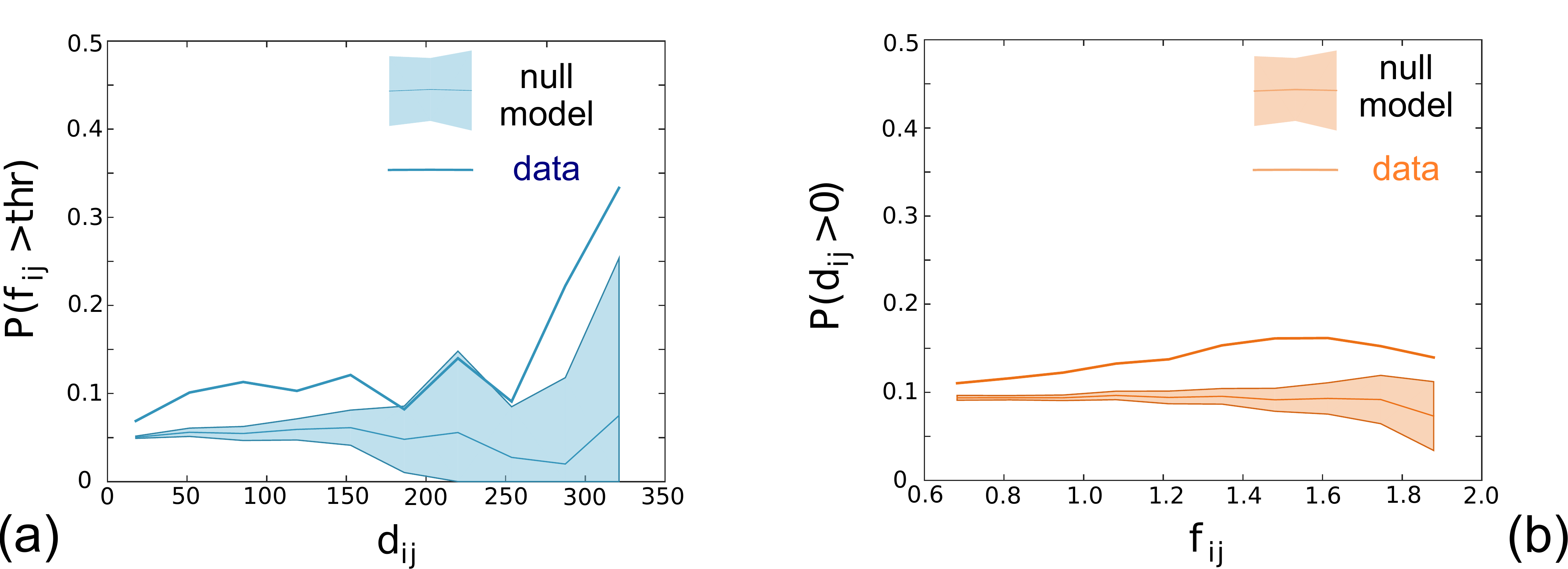}
\caption[]{Probability to observe a positive functional link given its weight on the structural layer (a) and to find an anatomical connection between two regions given the intensity of the correlation of their activity (b). In both cases data shows positive trends which are not observed in the corresponding null-models.}
\label{fig3}
\end{center}
\end{figure*}

For a more in-depth analysis of the relationship between functional and structural connections, we measured the probability $P(f_{ij}>\rm{thr})$ that there exists a significant positive correlation between two ROIs in the functional layer as a function of the weight $d_{ij}$ of their connection in the structural layer [Figure~\ref{fig3} (a)], and the dual probability $P(d_{ij}>0)$ that a structural link between two ROIs exists as a function of the strength of the correlation of their fMRI activity[Figure~\ref{fig3} (b)], to investigate the existence of network reinforcement mechanisms~\cite{battiston14} in the human brain. In such analysis we focus only on significant functional links which are positive, since the motif -Y was shown to be non-significant.

We show how in real data stronger values of physical connectivity are typically associated with a higher probability to have significant positive functional activity. A positive, but non-monotonic, trend also displays when plotting the probability to have a structural link given the weight of the functional correlation between the same two regions. In agreement with previous results~\cite{deco2011, honney2007, honney2009, honney2010}, our findings suggest that anatomical connectivity could predict well functional interactions across most of the brain areas. In a lesser degree, structural connections could also predict resting state connectivity. In both cases such positive trend is not observed in the corresponding null-models. In the case of Figure~\ref{fig3} (a), the null-model keeps fixed the weighted structure of the $\mathcal D$ and applies a block configuration model to the positive functional links of $\mathcal F$. Conversely, in Figure~\ref{fig3} (b), the structure of the weighted positive functional links is preserved and the binary structure of $\mathcal D$ is randomised through a block configuration model. 

\section{Conclusions}

The recent prevalence of applications involving multidimensional and
multimodal brain data has increased the demand for technical
developments in the analysis of such complex data. Modeling the human
brain as a complex network has unveiled the presence of charactaristic
non-trivial connectivity features (small-worldness, power-law degree
distributions and modularity among others) in many structural and
functional networks~\cite{bullmore2009}. The interplay between
anatomical connections and functional interactions is a current
challenge for understanding general brain functioning. In the last decade, some studies have thus
directly compared these connectivity structures to better investigate
possible direct mappings at the network level~\cite{adachi2012,
  simas2015}. In this study we addressed a fundamental problem in
multimodal brain networks analysis: the organisation of the complex
mosaic of brain motifs in anatomical and functional connectivity.

By considering multi-layer networks we have identified nonrandom motif
structures in multimodal brain networks. In contrast to current
approaches, which considers motif analysis on separate brain
modalities~\cite{sporns04, sakata2005, iturria2008}, this work
provides the first evidence of joint anatomo-functional motifs in
human brain networks. In line with previous
studies~\cite{damoiseaux2009, baria2011}, our results confirm the
complex relationships between structural connectivity and coupling of
brain dynamics. Significant multi-layer triads differ from those
usually obtained from single structural connectivity in both
humans~\cite{iturria2008} or monkeys~\cite{sporns04, adachi2012}. This
suggests that the multi-layer formalism constitutes could be a very
appropriate choice for the analysis of multimodal brain networks.

Our finding of a positive fMRI correlation between brain areas connected by a direct physical link is consistent with prior works~\cite{skudlarski2008}. In some cases, this increased functional connectivity can be explained by the spatial proximity of areas~\cite{bullmore2009}, but distant regions can also display strongly coherent dynamics without direct physical connections~\cite{nicosia2014}. In agreement with previous works, over-represented multi-layer triads involving negative functional links could also suggest a decrease in the anatomical connectivity that correlates with negative resting state correlations~\cite{skudlarski2008}. Nevertheless, more refined neuro-computational models are needed to fully explain the mechanism of this phenomenon.

A limitation of our study is the symmetrical configuration of the analyzed brain connectivity matrices, which is a consequence of the inherent symmetrical properties of DW- MRI techniques (fiber estimation cannot distinguish between afferent and efferent projections) and of the undirected functional interaction obtained by linear correlations (for a network of $N=264$ nodes, directed interactions could be estimated but from much longer time series~\cite{schlogelBook}). The symmetrical property of structural interactions constrains thus the motif analysis to consider a reduced number of motifs.

To conclude, our results indicate that the functional coordination of human brain dynamics at rest is non-trivially constrained by its underlying anatomical network, and suggest that structural connections might be necessary but not sufficient for the existence of positive functional correlations between two regions of the brain. Although we cannot definitively provide a one-to-one mapping of the structural and functional connectivity, we think that an anatomo-fonctional description of brain motifs might provide, more in general, meaningful insights into the organization of brain networks and the neural information processing during diverse cognitive or pathological states. We therefore hope that our approach will foster more principled and successful analysis of multimodal brain connectivity datasets.

\section{Acknowledgments}
This work was partially supported by the EU-LASAGNE Project, Contract No.318132 (STREP). 
F.B. thanks Nicholas Day and Trevor Pinto for useful suggestions.

\end{document}